\documentclass[aps,prb,twocolumn,showpacs]{revtex4}
\usepackage{bm}
\usepackage{graphicx}
\usepackage{amsmath}
\usepackage{eufrak}
\usepackage{color}
\newcommand{\nix}[1]{}

\begin{document}

\title{Nonlinear magneto-gyrotropic photogalvanic effect}
\author{H.~Diehl,$^1$  V.A.~Shalygin,$^2$ L.E.~Golub,$^3$ S.A.~Tarasenko,$^3$ S.N.~Danilov,$^1$ 
V.V.~Bel'kov,$^{1,3}$ E.G.~Novik,$^4$ H.~Buhmann,$^4$ C.~Br\"{u}ne,$^4$
 E.L.~Ivchenko,$^{3}$ and S.D.~Ganichev$^{1}$
}
\affiliation{$^1$Terahertz Center, University of Regensburg, 93040 Regensburg, Germany}

\affiliation{$^2$St.\,Petersburg State Polytechnic University, 195251 St.\,Petersburg, Russia}
\affiliation{$^3$A.F.\,Ioffe Physical-Technical Institute of the Russian Academy of Sciences, 194021 St.\,Petersburg, Russia}
\affiliation{$^4$ Physical Institute (EP3), University of
W\"{u}rzburg, 97074 W\"{u}rzburg, Germany}

%\date{\today}

\begin{abstract}
We report on the observation of nonlinear magneto-gyrotropic photogalvanic effect
in HgTe/HgCdTe quantum wells.
The interband absorption of  mid-infrared 
radiation as well as the intrasubband absorption of terahertz radiation 
in the heterostructures
is shown to cause a dc electric current in the presence of an in-plane
magnetic field. 
A cubic in magnetic field component of the photocurrent is observed 
in  quantum wells with the inverted band structure only.
The  experimental data are discussed in terms
of both the phenomenological theory and microscopic models.
\end{abstract}
% 73.50.Pz Photoconduction and photovoltaic effects
% 72.25.Fe Optical creation of spin polarized carriers
% 72.25.Rb Spin relaxation and scattering
% 78.67.De Quantum wells

\maketitle

\section{Introduction}

Much current attention in condensed matter physics is directed
toward understanding the spin dependent phenomena, both from the
fundamental point of view and due to increasing interest in
spintronics devices that are based not only on the electron
charge, but also on its spin. Conventional quantum well (QW) structures fabricated of III-V 
and II-VI wide gap materials are in focus of present
day  investigations. QW structures based on HgTe
appear to be  very attractive for the study of fundamental
spin-orbit effects. Narrow gap HgTe-based QWs are characterized by
an extraordinary large Rashba-type spin-orbit splitting, a
parameter crucial for the field of spintronics because it allows
an electric field control of spins, determines the spin relaxation
rate, and can be utilized for all-electric spin
injection.\cite{x} The lifting of spin degeneracy is caused by
spin-orbit interaction due to structure and bulk inversion
asymmetries which lead to Rashba and Dresselhaus spin-orbit terms
in the Hamiltonian, respectively (see Refs.~\onlinecite{Bychkov84p78,review2003spin,A1Zawadzki2003pR1,Winkler07,Fabian07,Dyakonov08}).
%\cite{Bychkov84p78,Dyakonov86p110,Lommer85p6985}
The Rashba spin splitting in HgTe-based QWs can reach values of up
to 30~meV, which is several times larger than for any other
semiconductor materials, and can be tuned over a wide
range.\cite{Hinz06,Gui04} Last but not least, HgTe-based QWs are
characterized by a highly specific band structure which, depending
on the well width and temperature, can be either normal or
inverted, small effective masses about
$0.02 \div 0.04 m_0$ (Refs.~\onlinecite{Pfeuffer98,Kvon2008}) and a large 
$g$-factor of about~20~(Ref.~\onlinecite{x2}). Despite the
enhanced spin features, however, there has been only a low
interest in the HgTe-based QWs. This can be attributed to
difficulties in the fabrication of HgTe-based devices and its
moderate mobilities. Recently, a significant progress has been
achieved in the growth of HgTe-based QWs. These advances make high
mobility samples available.\cite{Becker07} Additionally,
lithographical techniques were developed which meet the special
requirements of HgTe QWs.\cite{Daumer03}

The appearance of high quality HgTe/HgCdTe QWs resulted in the
observation of numerous transport, optical and magneto-optical
spin-related effects, like large Zeeman spin splitting,\cite{x3}
circular photogalvanic
effect,\cite{CPGEHgTe,CPGEHgTe2} enhancement of the subband spin
splitting by introducing magnetic ions in the QW structure\cite{Liu03,Gui04E} 
and the quantum spin Hall
effect.\cite{Koenig07,Koenig2008} The latter effect is
characterized by nondissipative transport of spin-polarized
electrons  and has a high potential for spintronics applications.

Here, we report on the observation of the magneto-gyrotropic
photogalvanic effect (MPGE) in (001)-grown HgTe/HgCdTe QWs. We
present  the  experimental and theoretical studies of MPGE induced
by terahertz as well as mid-infrared radiation.
The effect was detected %in (001)-oriented QWs 
in a wide temperature range   from liquid helium to 
room temperature. The MPGE has so far been detected in
GaAs, InAs, GaN, and Si QWs for various spectral
ranges~(for a review see Ref.~\onlinecite{Belkov08}). It has been shown
that different microscopic mechanisms of both paramagnetic\cite{Belkov05,Ganichev06}
(spin dependent) and
diamagnetic\cite{Gorbatsevich93,Kibis99,Diehl07,Tarasenko08}
origins can contribute to the photocurrent.
Recently, we demonstrated that MPGE provides a tool to probe the
symmetry of QWs and gives the necessary feedback to reliable
growth of structures with the controllable strength and sign of
the structure inversion asymmetry.\cite{PRL2008} Thus, the
observation of MPGE gives new access to
the novel material under investigation.
According to the previous studies carried out on III-V-based heterostructures 
the MPGE current depends
linearly on the magnetic field strength $B$. To our surprise, 
in HgTe/HgCdTe QWs with inverted band
structure we have detected both linear and nonlinear-in-$B$ contributions.
By contrast, in QWs with the normal band ordering the nonlinear-in-$B$ 
photocurrent is negligibly small.
The paper is organized as following. In Sec.~\ref{samples} we
give a short overview of the experimental technique. In
Sec.~\ref{results} the experimental results are summarized. In
Sec.~\ref{phenomenology} we  present the phenomenological
theory of the MPGE and compare its results with experimental data
on polarization dependences.
In Sec.~\ref{bandoptical} and~\ref{models} we
show the results of the band structure calculations and
discuss experimental data in view of the microscopic background.

\begin{figure}[b]
\includegraphics[width=0.75\linewidth]{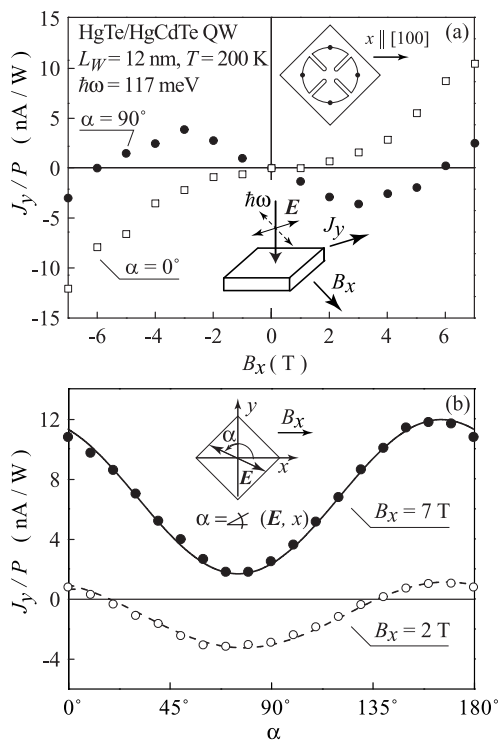}
%\centerline{\epsfxsize=0.85\linewidth \epsfbox{fig05.eps}}
%
\caption{Magnetic field and polarization dependences of the photocurrent 
measured in a QW structure with the well width $L_W=12$~nm
at temperature 200~K. Data are presented for normally incident mid-infrared radiation
with $\hbar \omega = 117$~meV ($\lambda = 10.57$~$\mu$m) and $P \approx 0.3$~kW. 
The magnetic field $\bm B$ is applied
parallel to the $x$ axis and the photocurrent is measured
in the direction $y$ normal to the vector $\bm B$.
(a) Magnetic field dependence for two states of polarization
with the azimuth angle $\alpha$ equal to $0^\circ$ and $90^\circ$.
(b) The dependence of the photocurrent on $\alpha$
 measured for two magnetic field strengths.
 The data on polarization dependence are fitted after Eqs.~(\ref{J0J0})
 and (\ref{j_exp}).
The insets show the experimental geometry and the orientation
of the light electric field ${\bm E}$ and the magnetic
field ${\bm B}$ with respect to the sample orientation.
} \label{fig2}
\end{figure}

\section{Samples and experimental technique}
\label{samples}

The experiments are carried out on
Hg$_{0.3}$Cd$_{0.7}$Te/HgTe/Hg$_{0.3}$Cd$_{0.7}$Te QWs having
four different nominal well widths, $L_W$: 5~nm, 8~nm, 12~nm
and 22~nm. Structures are molecular beam epitaxy (MBE) grown on a
Cd$_{0.96}$Zn$_{0.04}$Te substrate with the surface orientation
(001). Samples with the sheet density of electrons $n_s$ from $1
\times 10^{11}$~cm$^{-2}$ to $2 \times 10^{12}$~cm$^{-2}$ and
mobility in the range between $5 \times 10^4$ and $2 \times 10^5$
cm$^2$/Vs at $T=4.2$~K have been studied.
In order to investigate photocurrents, we have fabricated clover-shaped mesa structures
{of 4~mm diameter}  (see inset in Fig.~\ref{fig2})
using electron beam lithography and dry-etching techniques.
Ohmic contacts are fabricated by thermal In-bonding.
The contacts in clover-structures  are oriented along the $x \parallel [1 0 0]$ and
$y \parallel [0 1 0]$ crystallographic directions.
The photocurrent is measured in unbiased
structures via the voltage drop across a $50\: \Omega$ load
resistor.
Samples were mounted in  an optical cryostat which allowed us to
study MPGE in the temperature range from 4.2~K up to room temperature.
 An external in-plane magnetic field $B$ up to $\pm 7$~T could be
applied in the $x$-direction using a superconducting magnet. 

\begin{figure}[t]
\includegraphics[width=0.98\linewidth]{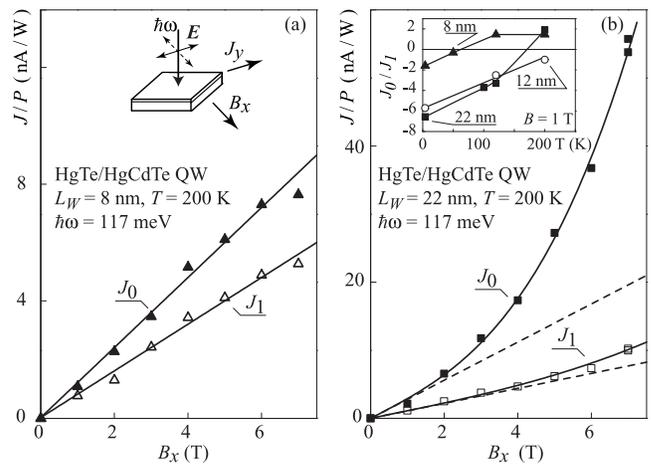}
%\centerline{\epsfxsize=0.85\linewidth \epsfbox{fig05.eps}}
%
\caption{Magnetic field dependence of  (a) the polarization
independent photocurrent $J_0$ and (b) the polarization dependent
photocurrent $J_1$ obtained for
the QW structures with $L_W= 8$~nm and 22~nm
at temperature 200~K and   
Data are given for normally incident
radiation of $P \approx 0.3$~kW
and the photon energy $\hbar \omega = 117$~meV. 
The data are fitted after Eqs.~(\ref{B1B3}) and (\ref{j_exp}). 
For the QW structure with $L_W= 8$~nm the fitting is limited by linear terms.
Dashed lines on the right panel demonstrate the linear contribution only.
Insets show the experimental geometry and the temperature
dependence of the ratio of polarization
independent and dependent photocurrents for QWs with $L_W= 8$~nm, 12~nm, 
and 22~nm at $B = 1$~T. 
}
\label{fig3}
\end{figure}

\begin{figure*}[t]
\includegraphics[width=0.8\linewidth]{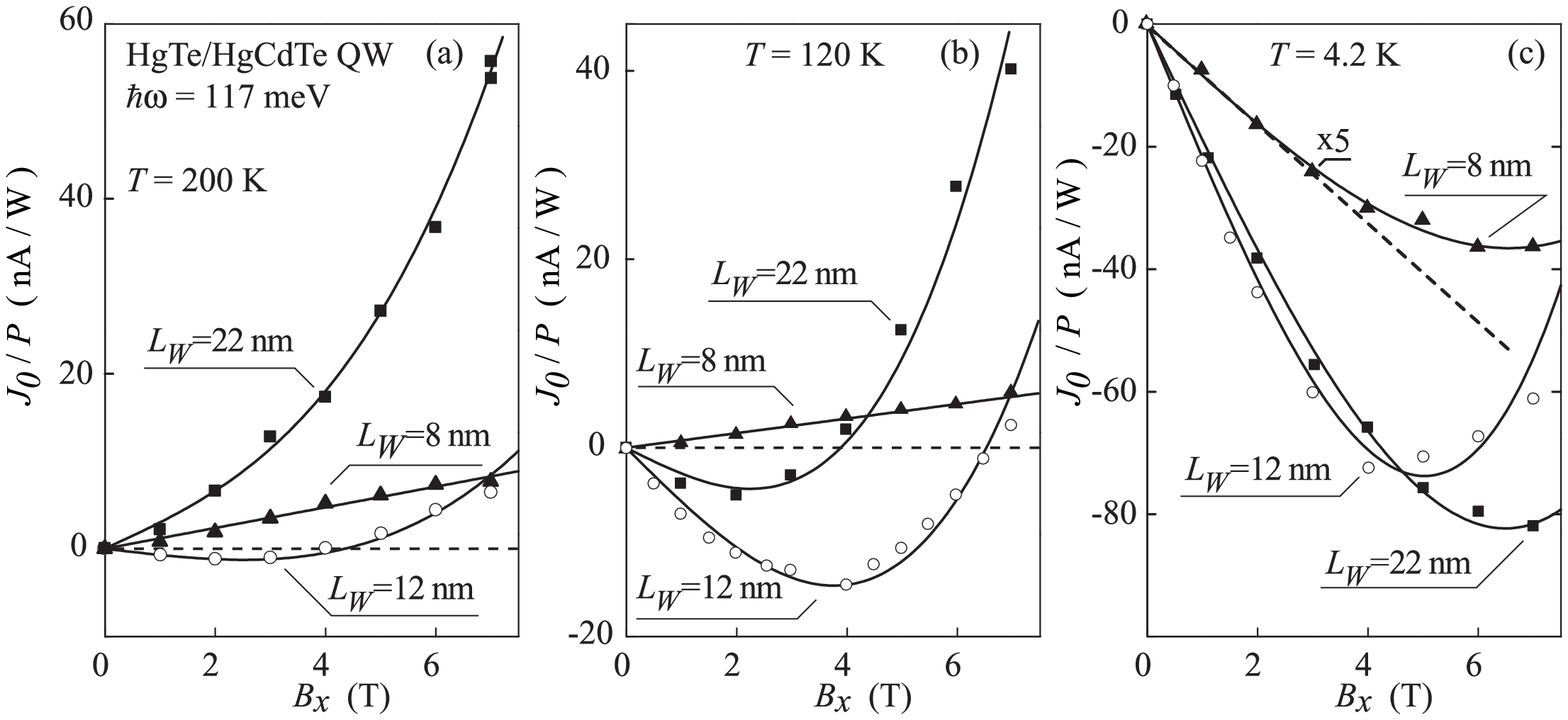}
\caption{Magnetic field dependences of the polarization independent
photocurrent $J_0$ obtained for QW
structures at  different temperatures.  
Data are given for normally
incident radiation of $P \approx 0.3$~kW and the photon energy
$\hbar \omega = 117$~meV. The
photocurrent is measured  in the direction  perpendicular to ${\bm
B}$ in QWs of three different widths. 
The data are fitted according to Eqs.~(\ref{B1B3}) and (\ref{j_exp}). 
The dashed line in the  panel (c) is plotted according to the linear law.
} \label{fig4}
\end{figure*}

The measurements of magnetic field induced photocurrents are
carried out under excitation of the samples with mid-infrared and
terahertz radiation at normal incidence.
The geometry of the experiment is sketched in the inset in
Fig.~\ref{fig2}(a). In (001)-oriented unbiased quantum well
structures this experimental arrangement excludes other effects
known to cause photocurrents.\cite{Ganichev06} The source of
infrared radiation is  a $Q$-switched CO$_2$ laser with operating
wavelengths $\lambda = 9.2 \div 10.8$~$\mu$m (corresponding photon
energies $\hbar \omega = 135 \div 115$~meV). In the investigated
narrow gap QWs the radiation of these photon energies may induce
$inter$-band optical transitions  or transitions between
size-quantized subbands. While the direct optical transitions
dominate in the radiation absorption, the less intensive free
carrier absorption (Drude-like) may contribute substantially to
the photocurrent generation.
The radiation power $P$ was varied in the range from 10~W up to 1.2~kW.
For the measurements in the terahertz range  we
used molecular laser, optically pumped by a TEA CO$_2$
laser.\cite{GanichevPrettl} With  NH$_3$ as active gas, 100~ns pulses of
linearly polarized
radiation with peak power $\sim$3~kW are obtained at  wavelengths
$\lambda= 90$, 148 and 280~$\mu$m (corresponding photon energies
$\hbar \omega$ are 13.7~meV, 8.4~meV and 4.4~meV). We also used a
CH$_3$F as active gas to obtain radiation with $\lambda=
496$~$\mu$m ($\hbar \omega = 2.5$~meV). The photon energies in the
terahertz range are smaller than the band gap as well as the
size-quantized subband separation and at moderate temperatures
terahertz radiation induces only free carrier absorption in the
lowest conduction subband. At low temperatures this radiation may
also cause direct transitions due to ionization of impurities, and
transitions between spin-split subbands due to Zeeman or Rashba
effects.

In our experiments we used the linearly polarized radiation.
In order to vary the angle $\alpha$ between the light
polarization plane and the magnetic field, the plane of
polarization of the radiation incident on the sample was rotated.
Hereafter the angle $\alpha = 0^\circ$ is chosen in such a way
that the incident light polarization is directed along the $x$
axis, see inset in Fig.~\ref{fig2}(b). In the terahertz range we
used $\lambda/2$-plates, which enabled us to change the azimuth
angle $\alpha$ from $0^\circ$ to $180^\circ$ covering all possible
orientations of the electric field vector in the QW plane.
In the mid-infrared range we applied a Fresnel rhomb converting
the linearly polarized laser radiation into the circularly
polarized radiation and placed an additional
double-Brewster-window polarizer behind the rhomb.
Rotation of the polarizer enabled us to tune the azimuth angle $\alpha$.

\section{Experimental results}
\label{results}

First, we discuss the results obtained with the mid-infrared
radiation. Irradiating  samples  at normal  incidence we observe,
for the in-plane magnetic field ${\bm B} \parallel x$, a
photocurrent signal in the $y$ direction. The width  of the
current pulses is about 300~ns  which corresponds to the infrared
laser pulse duration. The signal linearly depends on the radiation
power up to $P \approx 1.2$~kW, the highest power used in our
mid-infrared experiments. In Fig.~\ref{fig2}(a) the magnetic field
dependence of the photocurrent is plotted for HgTe/HgCdTe QW
structure with the well width of 12~nm. The data are obtained at
$T=200$~K for two polarization states of the radiation with the
electric field ${\bm E}$ of the light wave aligned parallel and
perpendicularly to the magnetic field. In the both cases the signal
is an odd function of $\bm B$. Its strength and behavior upon
variation of $B$ depends, however, on the orientation of the
radiation electric field vector. Figure~\ref{fig2}(b)  shows the
dependence of the photocurrent $J_y$ on the orientation of
polarization plane specified by the angle $\alpha$. The data can
be well fitted by the equation
\begin{equation}
\label{J0J0}
J_y(\alpha, B_x)  = J_0(B_x) + J_1(B_x) \cos{2\alpha} + J_2(B_x) \sin{2\alpha}.
\end{equation}
Below we demonstrate that exactly these
dependences follow from the theory.
The measurements in the two fixed polarization directions allow us
to extract {two} individual contributions:\cite{footnote1}
the polarization independent background and the amplitude of
one of the polarization dependent contributions, namely,
\begin{equation}
\label{J0J1}
J_0=
\frac{J_y(0^\circ)+J_y(90^\circ)}{2}\,,\,\,\,\,\,\,%\,\,\,\,
J_1= \frac{J_y(0^\circ)-J_y(90^\circ)}{2}\,\,.
\end{equation}
Figure~\ref{fig3} shows magnetic field dependence of~$J_0$ and~$J_1$
for samples with the well widths of 8~nm and 22~nm at $T=200$~K.
The signal behavior is different for these structures. We
have found that, for the QW with $L_W = 8$~nm, the photocurrent
depends linearly on the magnetic field.
On the other hand, in the QW with $L_W=22$~nm the
photocurrent can be described by a superposition of linear-in-$B$
and cubic-in-$B$ terms: 
\begin{equation}
\label{B1B3}
J_y(B) = a B + b B^3\,\,.
\end{equation}
Figure~\ref{fig3} shows that the $B^3$-term is  more pronounced in
the polarization independent photocurrent~$J_0$.
We focus below particularly on this
photocurrent because our measurements
reveal that this contribution dominates the photocurrent in the almost whole
temperature range even at low magnetic fields,
where the total photocurrent is mostly linear in $B$.
While the linear dependence of the photocurrent on magnetic field is
previously reported for various structures the observation of the cubic
in magnetic field photocurrent is unexpected and
has not been detected so far.
We emphasize that the last term in Eq.~\eqref{B1B3}
corresponding to $J_0$ is strong and overcomes the linear-in-$B$
contribution at the magnetic field about 6~T.

\begin{figure}[t]
\includegraphics[width=0.85\linewidth]{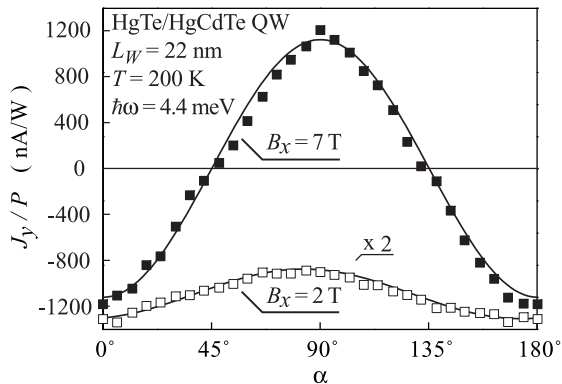}
%\centerline{\epsfxsize=0.85\linewidth \epsfbox{fig05.eps}}
%
\caption{Polarization dependence of the photocurrent $J_y$ excited
by terahertz radiation in the QW structure
with $L_W=22$~nm. The dependence is obtained at $T=200$~K, photon
energy $\hbar \omega = 4.4$~meV ($\lambda = 280\:\mu$m), radiation
power $P \approx 50$~W and for two magnetic field strengths.
The full lines are the fits after Eqs.~(\ref{J0J0}) and (\ref{j_exp}). }
\label{fig5Nalpha}
\end{figure}

Similar behavior was observed in the structure with $L_W = 12$~nm.
Moreover, in this sample
the coefficients $a$ and $b$ for polarization independent photocurrent~$J_0$ have
opposite signs resulting in a sign inversion observed for~$B$ about~4~T [see Fig.~\ref{fig4}(a)].
In the structure with $L_W = 5$~nm the signals were too small to
conclude definitely on the magnetic field dependence (but it is measurable 
at the excitation with THz radiation).
The decrease in temperature drastically affects the experimental
data. At intermediate temperature of 120~K we have observed that
the linear-in-$B$ contribution in QW with $L_W=22$~nm changes its
sign [see Fig.~\ref{fig4}(b)]. Now, the sample with $L_W=22$~nm also
shows the sign inversion of the photocurrent $J_0$ with rising
$B$, in the first sample with $L_W=8$~nm the data are still well
described by the linear-in-$B$ dependence. Further reduction of
temperature to the liquid helium temperature results in the sign
inversion of the linear-in-$B$ current in sample with $L_W=8$~nm
but also yields to the cubic-in-$B$ component [see
Fig.~\ref{fig4}(c)]. Now, the magnetic field dependence of the
photocurrent in all samples is described by the linear- and
cubic-in-$B$ terms with pre-factors of opposite signs. The total
current tends to the sign inversion, however at substantially
larger magnetic fields~$B$.

\begin{figure}[t]
\includegraphics[width=0.85\linewidth]{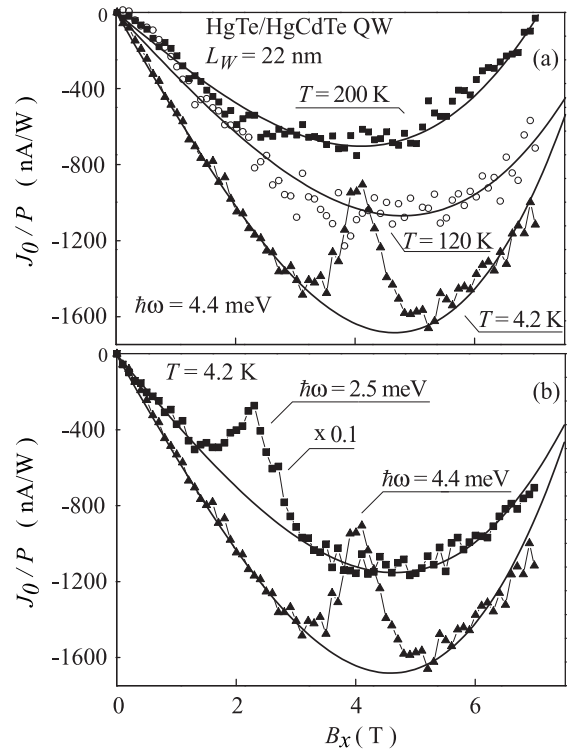}
%\centerline{\epsfxsize=0.85\linewidth \epsfbox{fig05.eps}}
%
\caption{Magnetic field dependence of the polarization independent
photocurrent $J_0$ excited by terahertz radiation in the QW structure 
with $L_W=22$~nm. (a) The photocurrent is
measured in the direction  perpendicular to ${\bm B}$ in response
to the radiation of $\lambda = 280\:\mu$m of $P \approx 50$~W
measured at three temperatures. (b) The photocurrent is measured at
liquid helium temperature in response to the radiation of two
photon energies.
The lines are plotted according to Eqs.~(\ref{B1B3}) and (\ref{j_exp}).
 } \label{fig5N}
\end{figure}

\begin{figure}[h]
\includegraphics[width=0.7\linewidth]{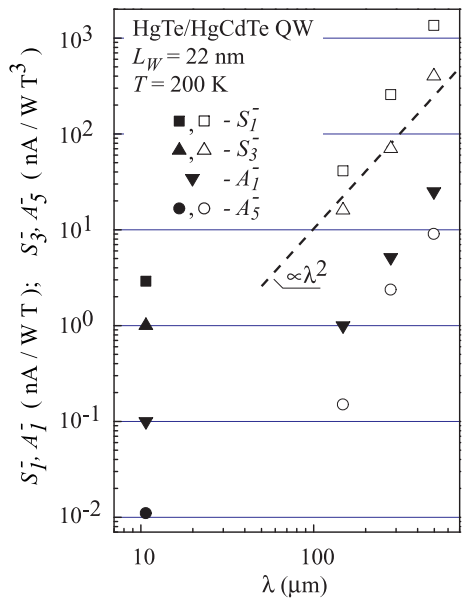}
%\centerline{\epsfxsize=0.85\linewidth \epsfbox{fig05.eps}}
%
\caption{Wavelength dependences of the absolute values of coefficients
$S_1^-,S_3^-,A_1^-$ and $A_5^-$ [see Eq.~(\ref{j_exp})] obtained
for the QW structure with $L_W=22$~nm at $T=200$~K. 
Full symbols correspond to negative values of the coefficients.
The dashed line is plotted according to the wavelength square law.
} \label{fig_lambda}
\end{figure}

Now we turn to the experiments with terahertz radiation. We
observed magnetic field induced photocurrent in all structures,
including sample with $L_W = 5$~nm and at all wavelengths used.
Like in the mid-infrared range the signal depends on the radiation
polarization (see Fig.~\ref{fig5Nalpha}) and is well described by
Eq.~(\ref{J0J0}). Figure~\ref{fig5N}(a) shows the magnetic field
dependence of the polarization independent contribution to the
photocurrent $J_0$  obtained in the wide QW with $L_W = 22$~nm
in response to the radiation of the photon energy $\hbar \omega =
4.4$~meV ($\lambda = 280\:\mu$m). Figure~\ref{fig5N}(a) demonstrates
that also in the terahertz range the photocurrent in the  QW with
$L_W = 22$~nm is well described by the Eq.~(\ref{B1B3}) with
significant contribution of the cubic-in-$B$ term at high
magnetic field. At low temperature we also detected a peak in
the magnetic field dependence (a dip for absolute value of the signal). 
The peak has minimum at $B \approx 4$~T and a halfwidth of about 0.75~T. 
Applying radiation of $496\:\mu$m
wavelength we obtained that the magnetic field position of the
peak  linearly scales with the photon energy
[Fig.~\ref{fig5N}(b)]. At shorter wavelength, e.g., with the photon
energy $\hbar \omega = 8.4$~meV ($\lambda = 148\:\mu$m), no peak
has been detected at $B \leq 7$~T. Similar behavior is also detected in the
polarization dependent contribution $J_1$, however the peak in
this contribution is much less pronounced. 
Figure~\ref{fig_lambda}
demonstrates that  linear-in-$B$ as well
as cubic-in-$B$ current contributions $J_0$ and $J_1$ drastically increase 
with increasing of the
wavelength. We see that at longest wavelength used ($\lambda = 496\:\mu$m) all current
contributions are more than two orders of magnitude larger than
that detected in the mid-infrared range. We note that some
contributions invert the sign with wavelength increasing.

\begin{figure}[h]
\includegraphics[width=0.85\linewidth]{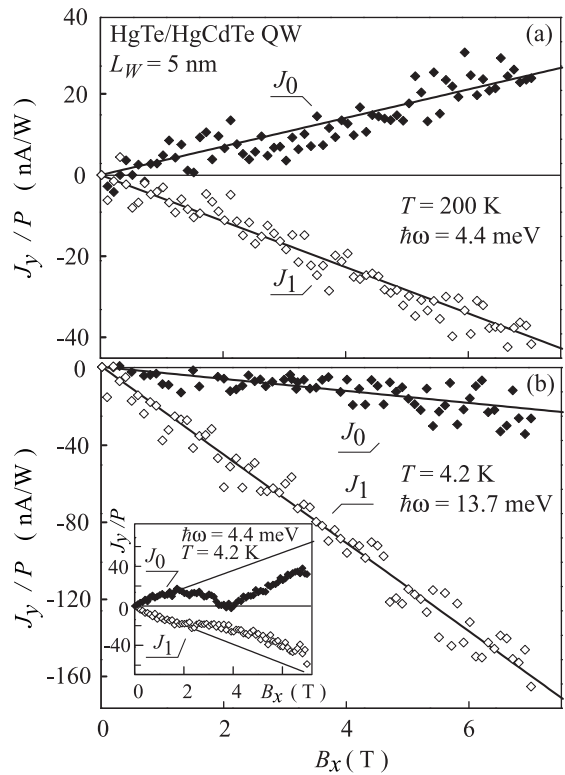}
%\centerline{\epsfxsize=0.85\linewidth \epsfbox{fig05.eps}}
%
\caption{Magnetic field dependence of the
photocurrent excited by terahertz radiation
in the QW structure with $L_W=5$~nm.
(a) The polarization independent and polarization dependent
contributions to the photocurrent measured at $T=200$~K in the
direction  perpendicular to ${\bm B}$ in response to the radiation
with the photon energy $\hbar \omega = 4.4$~meV ($\lambda =
280\:\mu$m) and $P \approx 3$~kW. (b) Photocurrent measured at
liquid helium temperature in response to the radiation with the
photon energy $\hbar \omega = 13.7$~meV ($\lambda = 90\:\mu$m).
The full lines are plotted according to Eqs.~(\ref{B1B3}) 
and (\ref{j_exp}) with coefficients $b$ and $A_{1,5}^-$ equal to zero.
The inset shows the data obtained at liquid helium temperature in
response to the radiation with $\hbar \omega = 4.4$~meV.
 }
\label{fig6N}
\end{figure}

In contrast to the wide QWs, in the narrowest QW sample ($L_W = 5$~nm) 
we observe that the photocurrent depends only linearly on the magnetic field~$B$. 
This is demonstrated in Fig.~\ref{fig6N}(a) for both,
polarization independent and polarization dependent, photocurrents
obtained for $T = 200$~K and excitation with the photon energy
$\hbar \omega = 4.4$~meV. The linear behavior of the photocurrent
is  observed even at low temperatures down to 4.2~K 
applying radiation of $\lambda = 90\:\mu$m ($\hbar \omega = 13.7$~meV) 
wavelength [Fig.~\ref{fig6N}(b)]. 
For a longer wavelength this behavior is masked by the wide peak  presented 
in the magnetic field
dependence of the photocurrent [see inset in Fig.~\ref{fig6N}(b)].
At the photon energy $\hbar \omega = 4.4$~meV the peak position is close to that
observed in the  QW with $L_W = 22$~nm, but it is much wider and is 
characterized by a halfwidth of at least  3~T. Like in the wide QWs, 
at higher photon energies no peak has been seen for $B \leq 7$~T
allowing one to analyze the magnetic field dependence unaffected by the peak.

\section{Phenomenology}
\label{phenomenology}

In order to describe the observed magnetic filed and polarization
dependences, we first derive here phenomenological equations for the
photocurrents in two-dimensional HgTe-based structures.
Holding the linear and cubic in the magnetic field strength $B$
terms, MPGE for unpolarized or linearly polarized radiation at
normal incidence is given by %%
\begin{eqnarray} \label{phen0}
j_\alpha &=& \sum_{\beta\gamma\delta}
\phi_{\alpha\beta\gamma\delta}\:B_\beta\:\frac{e_\gamma e^*_\delta
+ e_\delta  e^*_\gamma}{2} I \\
&+&  \sum_{\beta\mu\nu\gamma\delta}
\Xi_{\alpha\beta\mu\nu\gamma\delta}\:B_\beta B_\mu B_\nu\:\frac{e_\gamma
e^*_\delta + e_\delta e^*_\gamma}{2} I\:. \nonumber
\end{eqnarray}
Here $\bm{\phi}$ and $\bm{\Xi}$ are  a fourth- and a sixth-rank
pseudo-tensors, respectively, being symmetric in the last two
indices, $e_{\gamma}$ are components of the unit vector of light
polarization, and $I$ is the light intensity. We note that while
in the theoretical consideration the current density $\bm{j}$ is
used, in the experiments the electric current $\bm{J}$ is measured
which is proportional to the current density $\bm{j}$.

We consider (001)-oriented HgTe-based QWs. Depending on the
equivalence or nonequivalence of the QW interfaces their symmetry
may belong to one of the point groups D$_{2d}$ or C$_{2v}$,
respectively. The present experiments have been carried out on the
asymmetric structures of C$_{2v}$ symmetry and, therefore, here we will focus on
these QWs only. For the C$_{2v}$ point group it is convenient to
write the components of the magneto-photocurrent in the coordinate
system with $x' \parallel [1\bar{1}0]$,  $y' \parallel [110]$, and
$z \parallel [001]$ being the growth direction. The advantage of
this system is that the in-plane axes $x'$, $y'$ lie in the
crystallographic planes (110) and ($1\bar{1}0$) which are the
mirror reflection planes containing the twofold axis C$_2 \parallel z$.

In QWs of C$_{2v}$  symmetry
class the tensors $\bm{\phi}$ and $\bm{\Xi}$ have, respectively, six
and twelve linearly independent components and in the system $x'$,
$y'$, $z$ for normal incidence of the linearly polarized or unpolarized light 
and the in-plane magnetic field Eq.~\eqref{phen0} is reduced to:
\begin{widetext}
\begin{eqnarray}\label{MPGE_C2v}
j_{x'} &=& I \: \left[ S_1 B_{y'} + S_2 B_{y'} (|e_{x'}|^2-|e_{y'}|^2) + S_3 B_{x'} (e_{x'} e_{y'}^* + e_{y'} e_{x'}^*)  \right] \nonumber \\
&+& I \: B_{y'} [A_1B^2 + A_2 (B_{x'}^2-B_{y'}^2)] \nonumber \\
&+& I \: B_{y'}[A_3B^2+A_4(B_{x'}^2-B_{y'}^2)](|e_{x'}|^2-|e_{y'}|^2) \nonumber  \\
&+&I \: B_{x'} [A_5B^2+A_6(B_{x'}^2-B_{y'}^2)](e_{x'} e_{y'}^* + e_{y'} e_{x'}^*) \:,  \\
j_{y'} &=& I \: \left[ S_1' B_{x'} + S_2' B_{x'}
(|e_{x'}|^2-|e_{y'}|^2) + S_3' B_{y'} (e_{x'} e_{y'}^* + e_{y'} e_{x'}^*)  \right] \nonumber \\
&+& I \: B_{x'} [A_1' B^2 + A_2' (B_{x'}^2-B_{y'}^2)] \nonumber \\
&+& I \: B_{x'}[A_3' B^2 +
A_4'(B_{x'}^2-B_{y'}^2)](|e_{x'}|^2-|e_{y'}|^2)
\nonumber \\
&+&I \: B_{y'} [A_5'B^2+A_6'(B_{x'}^2-B_{y'}^2)](e_{x'} e_{y'}^* +
e_{y'} e_{x'}^*)\:. \nonumber
\end{eqnarray}
\end{widetext}
Here $S_i$ and $A_j$ are the linearly independent components
of the tensors $\bm{\phi}$ and $\bm{\Xi}$, respectively.
The polarization dependence of the photocurrent is determined by the factors  
($|e_{x'}|^2-|e_{y'}|^2$) and ($e_{x'} e_{y'}^* + e_{y'} e_{x'}^*$).

In our experiments the magnetic field was oriented along the cubic
axis ${\bm B} \parallel x$ and the current $J_{y}$
was measured perpendicularly to $\bm B$.
For this experimental geometry Eqs.~\eqref{MPGE_C2v} reduce to
\begin{eqnarray}
\label{j_exp}
    j_{y} = I B_x (-S_1^- + S_2^- \sin{2\alpha} - S_3^- \cos{2\alpha}) \\
    + I B_x^3 (-A_1^- + A_3^- \sin{2\alpha} - A_5^- \cos{2\alpha}), \nonumber
\end{eqnarray}
where $S_l^- = (S_l-S_l')/2$, $A_l^- = (A_l-A_l')/2$, and $\alpha$
is an angle between the linear polarization direction and  the
axis $x \parallel [100]$, see inset to Fig.~\ref{fig2}(b). {Thus,
for the polarization independent and polarization dependent
contributions to the photocurrent measured in the experiment we
have $J_0 \propto - (B_x S_1^- +  B_x^3 A_1^-)$, $J_1 \propto
-(B_x S_3^- + B_x^3 A_5^-)$ and $J_2 \propto (B_x S_2^- +   B_x^3
A_3^-)$. 

Equation~(\ref{j_exp}) describes well the macroscopic features of
the photocurrent. In accordance with the experimental data it
contains both linear- and cubic-in-$B$ contributions and fully
describes the observed polarization dependence (see
Figs.~\ref{fig2}(b) and~\ref{fig5Nalpha}]. Figure~\ref{fig3} shows that in the 
field $B \le 1$~T the linear part is the dominant one and it reveals both
polarization independent $J_0$ and polarization dependent $J_1$ parts.
According to Eq.~\eqref{j_exp} they are given by the coefficients
$S_1^-$ and $S_{3}^-$, respectively. The temperature dependence
of the ratio $J_0/J_1=S_1^-/S_3^-$   is presented in the inset to
Fig.~\ref{fig3}(b) and shows that polarization independent
contribution dominates the total photocurrent over almost all the
temperature range.
 In the narrowest QW with $L_W = 5$~nm and in QW with $L_W = 8$~nm
at high temperature the linear-in-$B$ behavior remains up to the highest 
magnetic fields applied.
In other samples, by contrast, for $B > 1$~T the cubic-in-B contribution
is clearly detected and even dominates the photocurrent.

\section{Band structure and optical transitions}
\label{bandoptical}

Now we calculate the band structure of our samples and indicate optical transitions
responsible for radiation absorption and the MPGE current generation.
HgTe as a bulk material is a zero-gap semimetal, whereas a narrow energy gap
opens up in a quantum well. Depending on the actual well width and temperature,
the band structure is either normal or inverted. In the latter case, the ordering of the subbands
in the QW is reversed compared to common semiconductors.

In Fig.~\ref{fig1a} the calculated band structure of 8~nm QW is
shown together with possible direct optical transitions
corresponding to the photon energy $\hbar \omega = 117$~meV used
in the experiment with mid-infrared radiation.
The band structure of (001)-grown HgTe/Hg$_{\scriptstyle
0.3}$Cd$_{\scriptstyle 0.7}$Te QW   was
calculated using the eight-band $\bm{k}$$\cdot$${\bm p}$ model in
envelope function approximation.\cite{Novik2005} This QW is a
type III heterostructure (see insets to Fig.~\ref{fig1a})
that causes mixing of the
electron states and strong coupling between the conduction and valence bands.
In order to take into
account the coupling and the resulting nonparabolicity of the
bands the Kane model with the usual eight-band
basis set $\{|u_{n0}\rangle\}=|\Gamma_6,\pm1/2\rangle,
|\Gamma_8,\pm1/2\rangle, |\Gamma_8,\pm3/2\rangle,
|\Gamma_7,\pm1/2\rangle$ was used. Assuming the basis functions
$u_{n0}$ to be the same throughout the heterostructure and using the
correct operator ordering in the effective-mass Hamiltonian for
the eight-component envelope function vector in accordance with
the envelope function theory\cite{Burt1999} the boundary
conditions at material interfaces are automatically satisfied. The
total eight-band Hamiltonian of the QW system is given by
$$H=H_0+H_1+H_2+V_H+H_{BP},$$
where $H_0$ is the
diagonal contribution including the band-edge potentials for
the chosen basis set $\{|u_{n0}\rangle\}$, $H_1$ and $H_2$
describe the coupling between the bands of this basis set exactly
and their coupling to the remote bands in second-order
perturbation theory, respectively, $V_H$ is the self-consistently
calculated Hartree-potential, and $H_{BP}$ is the Bir-Pikus
Hamiltonian describing the effects of strain in the structure.
The explicit form of the Hamiltonian as well as the band structure
parameters employed in the calculations are given in
Ref.~\onlinecite{Novik2005}. The obtained system of eight coupled
differential equations of the second order for the envelope
function components was transformed then into a matrix eigenvalue
problem by means of the expansion of the envelope function
components in terms of the complete basis set which results in the
required convergence for type~III heterostructures. The subbands
in Fig.~\ref{fig1a} are labeled as heavy-hole- ($H\textit{i}$),
electron- ($E\textit{i}$) and light-hole-like ($L\textit{i}$) in
accordance with the properties of the corresponding wave functions
at $k_\parallel=0$~(see Ref.~\onlinecite{Truchsess1998}).

\begin{figure}%[width=4cm]
\includegraphics[width=0.99\linewidth]{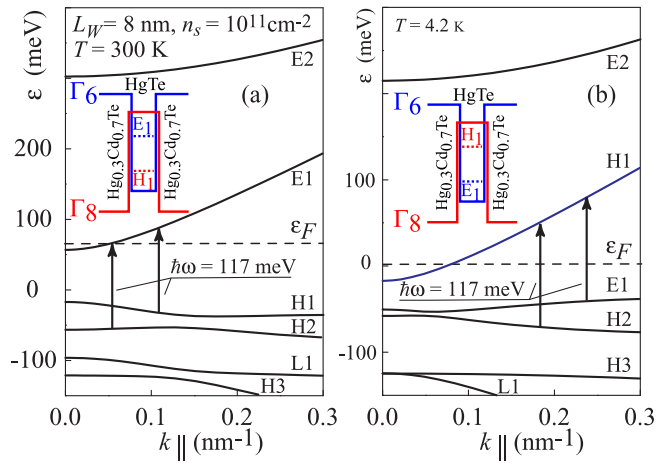}
%\centerline{\epsfxsize 70mm \epsfbox{fig1_structures.eps}}
\caption{Calculated band structure for 8~nm QW at (a) 300~K  and (b) 4.2~K.
Arrows show optical transitions induced by  mid-infrared radiation used
in the experiments ($\hbar \omega = 117$~meV). 
Insets sketch the band profile of noninverted  (left panel) and inverted (right panel)
HgTe-based QWs.
}
\label{fig1a}
\end{figure}

We emphasize that HgTe-based QW may have a
normal or inverted band structure depending on its width 
and the temperature.\cite{Koenig2008} For example, for $T=4.2$~K and QW
width $L_W\lesssim 6$~nm electron-like (hole-like) subbands form
the conduction (valence) band. Figure~\ref{fig1a} shows that while
at 4.2~K heterostructures with 8~nm QWs are characterized by the
inverted band structure, at room temperature it has a normal band
structure.
At $T=4.2$~K for $L_W \gtrsim 6$~nm H1 subband lies above the E1
subband and becomes the lowest conduction subband [see
Fig.~\ref{fig1a}(b)]. With temperature increasing  the critical width
is shifted to larger values and for $T=300$~K and $L_W =8$~nm the
QW has a normal sequence of the subbands. Calculations of the band
structure for QWs with $L_W =12$~nm and 22~nm 
demonstrate that they have an inverted band structure for all
temperatures used in our experiments, whereas QW of 5~nm width has 
noninverted structure in the whole temperature range. 
 The analysis of the
band structure of investigated samples reveals that the nonlinear
behavior of the MPGE is detected only in samples having inverted
band structure.

Our calculations show that mid-infrared radiation with the photon
energy of the order of 100~meV used in experiments causes in all
our samples direct interband optical transitions  (see
Fig.~\ref{fig1a}). The photon energies of applied terahertz radiation
($\hbar\omega =3 \div 14$~meV) are much smaller than the energy
gap and intersubband separation, therefore this radiation causes
only indirect (Drude-like) optical transitions. 
At low temperatures with $k_{\rm B}T < \hbar\omega$ terahertz radiation
may also cause ionization of impurities, intra-impurity  transitions 
or direct transitions between the Zeeman spin-split subbands.
These mechanisms may have a resonance-like behavior 
and be responsible for peaks observed in the magnetic field dependences
of the photocurrent at liquid helium temperature. A comparatively
 large width of these
peaks covering several Tesla indicates that they are most probably
due to  impurity related mechanisms. 
The magnetic field shifts the band edge as well as the
impurity level and tune the binding energy to the photon energy
making the direct optical excitation possible. The mechanism of this
additional channel of the radiation absorption and the resulting
MPGE are out of scope of this paper.

\begin{figure}%[width=4cm]
\includegraphics[width=0.99\linewidth]{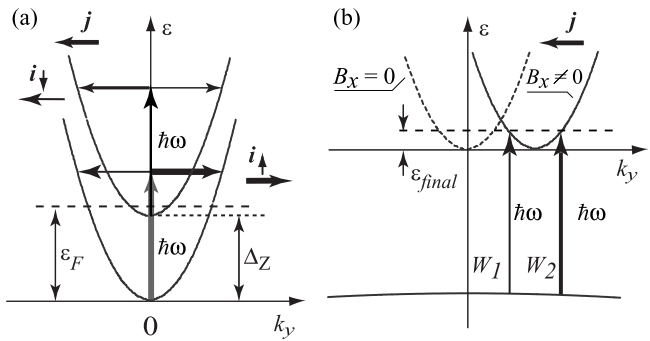}
%\centerline{\epsfxsize 70mm \epsfbox{fig1_structures.eps}}
\caption{Microscopic models  of MPGE (a) due to
imbalance of the spin photocurrents in the in-plane magnetic field
and (b) due to the diamagnetic mechanism at direct intersubband transitions.
}
\label{figmodel}
\end{figure}

\section{Microscopic models and discussion}
\label{models}

The most surprising result obtained in the experiment is that in samples
with $L_W=12$~nm and
 $L_W=22$~nm, as well as in the sample with $L_W=8$~nm at low temperature,
the cubic-in-$B$ contribution to $J$ is strong and may overcome
the linear-in-$B$ contribution at the magnetic field of 4$\div$5~T. 
Therefore, we focus below on possible microscopic mechanisms of
such a non-linear behavior
 which is observed in QWs with the inverted band structures only.
Since to the best of our knowledge the band structure of
HgTe/HgCdTe QWs in in-plane magnetic fields is not available, we
perform here only a qualitative microscopic analysis  of the
effect.

First we discuss the terahertz spectral range where radiation
absorption is dominated by Drude-like processes. In this case, the
photocurrent is mainly caused by asymmetry of the electron
scattering by phonons and static defects in the magnetic
field.\cite{Belkov08} Such a magnetic-field induced scattering
asymmetry can be of both spin dependent and diamagnetic
(spin independent) origins. The spin dependent mechanism of MPGE
comes from the imbalance of the spin photocurrents in the in-plane
magnetic field.\cite{Ganichev06} Microscopically, it is based on
spin dependent scattering which accompanies the free-carrier
absorption. Figure~\ref{figmodel}(a) sketches the indirect optical
transitions within two spin subbands. Vertical arrows indicate
optical transitions from the initial state $k_y = 0$ while the
horizontal arrows describe a scattering event to a final state
with either positive or negative electron wave vector $k^\prime_y$.
Due to the spin dependence of scattering the transitions to states
with positive and negative $k^\prime_y$ occur with unequal
probabilities. This is indicated by horizontal arrows of different
thicknesses. By that the free carrier absorption leads to a pure
spin current where particles with opposite spin orientations flow
in opposite directions.
Similarly to the excitation mechanism, energy
relaxation of electron gas heated by Drude absorption, also involving
electron scattering, is asymmetric and yields spin separation as well.
By application of an external magnetic field which polarizes free
carriers, the spin photocurrent is converted into an electric
current proportional to the Zeeman splitting for small fields. We
note that the mechanism based on asymmetry of the photexcitation
yields polarization dependent photocurrent while that related to
asymmetry of energy relaxation results in polarization independent
signal.\cite{Ganichev06} In QWs with inverted band structure
the ground conduction subband, which is populated in equilibrium, is
formed from the $\Gamma_8$-band states [see Fig.~\ref{fig1a}(b)]. The
Zeeman splitting $\Delta_{\rm Z}$ of heavy-hole states in the
in-plane magnetic field depends strongly nonlinear on
$B$~(Ref.~\onlinecite{Winkler_book}). Since for this mechanism  $j({\bm B})
\propto \Delta_{\rm Z}({\bm B})$, the photocurrent exhibits a
nonlinear behavior in the magnetic field.
In  the narrow QW with $L_W = 5$~nm the ground conduction subband
is formed from the $\Gamma_6$-band states. Here the Zeeman
splitting is linear in~$B$, and a noticeable cubic-in-$B$
contribution to the photocurrent is absent as observed in our
experiments.

The diamagnetic mechanism of the MPGE under free-carrier
absorption is also related to asymmetry of electron scattering in
the magnetic field.\cite{Kibis99,Tarasenko08} However the
asymmetry stems from the magnetic field induced mixture of states
from different quantum subbands, which is not related to the
Zeeman splitting. This  mixture is more efficient for subbands
formed from the same Bloch states (i.e., between $E1$ and $E2$ or
$H1$ and $H2$) and determined by the ratio of $\hbar e B/(m^* c)$
to the intersubband energy separation $\Delta \varepsilon$, where
$m^*$ is the in-plane effective mass. In QWs with inverted band
structure the ground $H1$ subband is close to the valence subbands
of hole type, see Fig.~\ref{fig1a}(b). Therefore in moderate
magnetic fields of several Tesla, the ratio becomes not small,
which leads to nonlinear dependence of the photocurrent on
magnetic field. By contrast, in the structure with normal band
arrangement, Fig.~\ref{fig1a}(a), the energy separation between the
ground subbands~$E1$ and $E2$ exceeds 200~meV, and the MPGE current
linearly depends on~$B$.

The mechanisms described above may also be responsible for the
photocurrent caused by mid-infrared radiation. Although the
contribution from the Drude processes to the total absorption does
not seem to be dominant in the spectral range where interband
transitions are possible, it may nevertheless determine the
photocurrent. This scenario is supported by the drastic spectral
dependence of the  photocurrent demonstrated in
Fig.~\ref{fig_lambda}. Indeed, the photocurrent strength increases
by more than an order of magnitude with increasing wavelength,
the dependence usually detected for Drude-like absorption.
Another contribution may come from the
direct intersubband transitions caused by
mid-infrared radiation.
Such a mechanism of MPGE is proposed in Ref.~\onlinecite{Gorbatsevich93} and
is based on magnetic field induced shift of quantum subbands in
$\bm{k}$-space which is described by the linear-in-$\bm B$
contribution to the electron energy given by $ \delta
E^{(\nu)}_{\bm k}= (e \hbar \,\bar{z}_{\nu} / c m^*) [\bm{B}\times
\bm{k}]_z,$ where
$\bar{z}_{\nu}$ is the mean coordinate along the growth direction,
and $\nu$ is the QW subband index. The energy spectrum of the QW,
including the diamagnetic shift, is sketched in
Fig.~\ref{figmodel}(b). In conventional QW structures with parabolic
dispersion of the valence and conduction subbands, the relative
subband shift leads to a photocurrent at direct optical
transitions.\cite{Gorbatsevich93,Diehl07} Indeed, in such systems
due to the energy and momentum conservation the points of optical
transitions are shifted in the $\bm{k}$-space resulting in
asymmetric distribution of photoexcited carriers with respect to
the subbands minima, i.e., to a magnetic field induced
photocurrent. However, this straightforward mechanism of MPGE gets
ineffective in HgTe-based structures under study where the valence
subbands are flat (see Fig.~\ref{fig1a}). In this particular case,
the points of optical transitions remain symmetric with respect to
the conduction subband minimum. Therefore, the diamagnetic shift
of the conduction subband does lead to a photocurrent only if the
probability of optical transitions $W$ depends on the wave vector.
Such a dependence may come from mixture of the states at finite
in-plane wave vector resulting in $W=w_0+w_2 k^2$. This term
together with the diamagnetic shift of the conduction subband give
rise to the photocurrent $j(B) \propto w_2 B$, see
Fig.~\ref{figmodel}(b). Moreover, in QWs with inverted band structure
and closely spaced valence subbands, like in the case of wide HgTe
QWs, one can expect that the parameter $w_2$ can be large enough
and the magnetic field has a remarkable effect on $w_2$. This
leads to a nonlinear dependence of the photocurrent on magnetic
field.

\section{Summary}

In conclusion, we have presented experimental data evincing the first observation of
nonlinear magnetic field dependence  of MPGE current. One of a probable scenarios
is based on the cubic-in-$B$ Zeeman splitting of the ground subband. 
To prove this statement additional experiments like electron spin resonance 
investigations are needed.
Further access to the origin of
the photocurrent and various mechanisms contributing in its formation should provide an analysis
of the temperature and spectral behavior, and, in particular, of the observed sign inversions.

\section*{ACKNOWLEDGMENTS}

We thank M.M.~Voronov for helpful discussions.  The financial support
%work is supported
of the DFG  via programs SFB~689, GA~501/6-3 and grant AS327/2-1
as well as support of RFBR is gratefully acknowledged.
Work of L.E.G. is also supported by ''Dynasty''
Foundation --- ICFPM and President grant for young scientists.

%\newpage

%%%%%%%%%%%%%%%%%%%%% Figure captions

%Figure 1


\begin{thebibliography}{99}
\bibitem{footnote1} We note that both polarization dependent components of the total photocurrent
in Eq.~(\protect\ref{J0J0}) stem from the same microscopic
mechanism, see Ref.~\protect \onlinecite{Belkov05}.
\end{thebibliography}
\end{document}